\documentclass[10pt,twocolumn,twoside,journal]{IEEEtran}
\IEEEoverridecommandlockouts
\usepackage{subfigure} %????
\usepackage{graphicx}

\usepackage{amsmath,graphicx,amssymb,mathtools,bm}
\usepackage{subfigure}
\usepackage{hyperref}
\usepackage{cite}
\usepackage{amsmath,amssymb,amsfonts}  
\usepackage{textcomp}
\usepackage{xcolor}
\usepackage{verbatim}  % \begin{comment} ... \end{comment} 包含被注释的文字 
\usepackage{bm}  % 数学环境下加粗字体$\bm{ a }$
\usepackage{mathrsfs} % 花体字母 
 \usepackage{algorithmic} %format of the algorithm
\usepackage{booktabs}
\usepackage{textcomp}  % 千分号\textperthousand 摄氏度：$^\circ$C \par
\usepackage{multirow}  % 多行合并
\usepackage{lettrine}   % 首字下沉
\usepackage{graphicx}  % 双栏模板插入通栏图片
\usepackage{color}  % 字体颜色
\usepackage{amsmath}
\usepackage{amssymb}
\usepackage{stfloats} 
\usepackage{caption} 

\usepackage{color}  % 字体颜色
\usepackage[ruled,linesnumbered]{algorithm2e}

\begin{document}
	\newcommand{\tabincell}[2]{\begin{tabular}{@{}#1@{}}#2\end{tabular}}
   \newtheorem{Property}{\it Property} 
  
 \newtheorem{Proposition}{\bf Proposition}
\newtheorem{remark}{Remark}
\newenvironment{Proof}{{\indent \it Proof:}}{\hfill $\blacksquare$\par}
%\makeatletter
%\newcommand\sixteen{\@setfontsize\sixteen{13pt}{6}}
%\renewcommand{\maketitle}{\bgroup\setlength{\parindent}{0pt}
%	\begin{flushleft}
%		\sixteen\bfseries \@title
%		\medskip
%	\end{flushleft}
%	\textit{\@author}
%	\egroup}
%\makeatother

\title{ Modeling and Optimization for Flexible Cylindrical Arrays-Enabled Wireless Communications
}
 
\author{
	Songjie Yang, Jiahe Guo, Zilin He, Boyu Ning, \IEEEmembership{Member,~IEEE}, Weidong Mei, \IEEEmembership{Member,~IEEE}, \\ Zhongpei Zhang, \IEEEmembership{Member,~IEEE}, Chadi Assi, \IEEEmembership{Fellow,~IEEE}, and Chau Yuen, \IEEEmembership{Fellow,~IEEE}

\thanks{
	Songjie Yang, Jiahe Guo, Boyu Ning,  Weidong Mei, and Zhongpei Zhang are with the National Key Laboratory of Wireless Communications, University of Electronic Science and Technology of China, Chengdu 611731, China. (e-mail:	yangsongjie@std.uestc.edu.cn; guojiahe@std.uestc.edu.cn; boydning@outlook.com; wmei@uestc.edu.cn; zhangzp@uestc.edu.cn).  
	 
	 Zilin He is with the School of Information and Control Engineering ,Jilin Institute of Chemical Technology (e-mail: nasarui7@outlook.com).
	 
	  Chadi Assi is with Concordia University, Montreal, Quebec, H3G 1M8,
	 Canada (e-mail:chadimassi@gmail.com).
	 
	 	Chau Yuen is with the School of Electrical and Electronics Engineering, Nanyang Technological University (e-mail: chau.yuen@ntu.edu.sg).
	 	}

}% <-this % stops a space

\maketitle

\begin{abstract}
	Flexible-geometry arrays have garnered much attention in wireless communications, which dynamically adjust wireless channels to improve the system performance. In this paper, we propose a novel flexible-geometry array for a $360^\circ$ coverage, named flxible cylindrical array (FCLA), comprised of multiple flexible circular arrays (FCAs). The elements in each FCA can revolve around the circle track to change their horizontal positions, and the FCAs can move along the vertical axis to change the elements' heights. Considering that  horizontal revolving can change the antenna orientation, we adopt both the omni-directional  and the directional antenna patterns. Based on the regularized zero-forcing (RZF) precoding scheme, we formulate a particular compressive sensing (CS) problem incorporating joint precoding and antenna position optimization, and propose two effective methods, namely FCLA-J and FCLA-A, to solve it. Specifically, the first method involves jointly optimizing the element's revolving angle, height, and precoding coefficient within a single CS framework. The second method decouples the CS problem into two subproblems by utilizing an alternative sparse optimization approach for the revolving angle and height, thereby reducing time complexity.  Simulation results reveal that, when utilizing directional radiation patterns, FCLA-J and FCLA-A achieve substantial performance improvements of 43.32\% and 25.42\%, respectively, compared to uniform cylindrical arrays (UCLAs) with RZF precoding.  

\end{abstract}
\section{Introduction}   
As wireless communication continues to evolve, the relentless push for innovation and technological advancements in various dimensions has led to a continuous stream of developments aimed at achieving higher degrees-of-freedom (DoFs) and expanding the scope of applications in this ever-growing field. One key focus in this evolution is the thorough exploration of the intrinsic characteristics of the wireless channel. Researchers and engineers seek to harness these unique properties to enhance both communication efficiency and sensing capabilities. 
At the forefront of these advancements is the exploitation of the inherent sparsity in millimeter-wave and terahertz frequency channels, which allows for advanced beamspace signal processing techniques \cite{mmw1,mmw2,NI}. Furthermore, the implementation of reconfigurable intelligent surfaces (RIS) has emerged as a groundbreaking method to significantly enhance channel propagation properties, as exemplified by research findings in \cite{RIS1,RIS2,M1}. RISs have the ability to dynamically alter the radio environment, providing a boost to both signal strength and coverage. 
Another exciting development lies in the exploitation of near-field spherical-wave channel characteristics, which opens up the possibility of unlocking an additional dimension in the form of the distance DoF \cite{NF1,NF2}. Despite these promising advancements, the full potential of wireless channels remains a vast and largely uncharted territory, offering an expansive field replete with opportunities for further explorative research and groundbreaking discoveries.

\subsection{Flexible-Geometry Arrays}

Recognizing the arrays' flexiblility, various researches across various fields exploit the array geometry's DoF to enhance the system performance. One classical example is array synthesis \cite{AAS1,AAS2}, which optimizes the antenna's excitation coefficient, position, and orientation to achieve the desired radiation pattern, such as high mainlobe and low sidelobe. Moreover, sparse array synthesis \cite{ars1,ars2} considers a more energy-efficient case by minimizing the number of antennas while ensuring the pattern synthesis performance. In this case, the antennas's positions have a large range, further emphsizing the significance of antenna position optimization. Usually, array synthesis is performed on a regular geometry, such as on a line, plane, circle, and cylinder. The flexible element position gives to irregular array geometry compared to the commonly used uniform arrays, enabling various DoFs to be utilized.

Several studies have analyzed the performance and signal processing of movable antennas (MAs) and fluid antenna systems (FAS) \cite{FA1, FA2, FA3, MA1, MA2, MA3}. These systems, at the first glance, are similar to the antenna selection topic \cite{AS1}, which involves choosing discrete antenna positions to adapt dynamically to channel conditions. However, beyond antenna selection, MAs/FAS do not require a large number of pre-deployed antennas and can achieve continuous optimization of antenna positions.
In \cite{MA1,M2}, the authors focused on maximizing the multi-path channel gain in MA systems, with continuous and discrete algorithms, respectively, which provide  valuable insights into MA-enhanced communications. Meanwhile, the authors of \cite{FA1} derived a closed-form expression for the lower bound of capacity in FAS, highlighting the significant capacity gains from the diversity inherent in a compact space. Both \cite{MA1,M2,FA1} theoretically illustrated the potential advantages of optimizing antenna positions in wireless communications.
The study in \cite{FA4} examined point-to-point FAS communications using maximum ratio combining, finding that the system's diversity order is equivalent to the total number of ports. Additionally, \cite{FA2} observed that in FAS, the multiplexing gain is directly proportional to the number of ports and inversely related to the signal-to-interference ratio target. The authors of \cite{MA2} and \cite{FA3} also demonstrated the possibility of minimizing uplink power by optimizing the user's antenna position. Conversely, \cite{MA3} investigated MAs at the base station (BS) using particle swarm optimization to maximize the minimum user rate by optimizing antenna positions. However, while such meta-heuristic algorithms are effective, they tend to have high computational complexity and may lack comprehensive theoretical insights. In \cite{MA_FP}, the authors introduced a sparse optimization framework called flexible precoding, which utilizes a linear precoding strategy to simultaneously optimize both antenna coefficients and positions. This approach aims to enhance the overall system performance. Furthermore, MAs have shown effectiveness when integrated with other emerging technologies. Notably, they have been successfully combined with integrated sensing and communications \cite{MAISAC1,MAISAC2},  cognitive radio \cite{M3}, and mobile edge computing \cite{MAMEC1,MAMEC2}. These integrations showcase the adaptability and potential of MAs in advancing various technological applications.

Building on the concept of flexible antennas \cite{FA1,FA2}, the flexible substrate enables the deformation of array structures, leading to the development of flexible antenna arrays \cite{FAA1, FAA2, FAA3, FAA4}. In particular, \cite{FAA2} examines the adaptability of linear antenna arrays when wrapped around a cylindrical surface, evaluating irradiation performance based on metrics such as side lobe levels and mutual coupling. The study in \cite{FAA1} explored a flexible array design capable of bending in both horizontal and vertical planes, with concave and convex radii of less than 23 cm, while retaining full functionality and programmability, including focusing, pattern recovery, and two-dimensional beam steering.
Beyond bending structures, folding structures are emerging as promising spaceborne devices in mechanical systems due to their unique transformable characteristics. Several studies on materials have explored the integration of folding into communication devices, achieving notable performance enhancements like wideband switchable absorption \cite{fold2}. Notably, \cite{fold1} employed folding techniques and flexible electronic materials in metamaterial design to achieve significant reflection modulation over an ultra-wideband spectrum, capitalizing on the folding degree of freedom. This approach holds potential for extensive satellite communications applications due to its lightweight, foldability, and cost-effective properties.
Additionally, \cite{FAA5} introduced a metasurface capable of adapting to various dynamic shapes through a Lorentz-force-driven mesh, which facilitates easier manipulation of shapes. This research highlights the potential of combining advanced materials and techniques to create intelligent and shape-adaptive systems.
These studies primarily focus on the antenna domain, examining FAAs from diverse perspectives, particularly in terms of radiation properties and structural adaptability.
Recently, \cite{P_FAA} expanded the concept of flexible arrays into the realm of wireless communications. By modifying the array's configuration in response to changes in the wireless channel, they aimed to enhance the multi-user sum-rate. Additionally, \cite{arrayMA} conducted an in-depth study of array-level MAs and their implementations, offering a promising avenue for the development of future MAs.

 \subsection{Cylindrical Arrays}
Despite existing work focusing on linear and planar arrays for wireless communications, circular and cylindrical arrays are also relevant in various scenarios. Notably, CLAs with their 3D geometry can achieve $360^\circ$
 beam coverage, making them more suitable for the massive access demands of 6G communications. In contrast, traditional 2D antenna arrays offer limited beam scanning, which fails to meet the 3D MIMO application requirements in terms of spatial DoF \cite{CLA4,CLA6}. The potential application of cylindrical massive MIMO arrays in cellular deployments, utilizing uniform beamforming gain in the azimuth/horizontal dimension, was explored in \cite{CLA7}. The study demonstrated that cylindrical arrays could provide higher minimum rates and less variance compared to sectorized UPAs, although some peak rate is sacrificed. This makes cylindrical arrays viable for applications requiring high reliability, such as industrial environments.
Moreover, the authors in \cite{CLA5} highlighted the potential of extremely large-scale CLAs to enhance near-field space division multiple access by designing a 3D near-field CLA codebook. In addition, \cite{CLA1} introduced a physical layer security method using massive CLAs, where some elements transmit signals reliably with directive beams, with others jamming eavesdroppers. The arrays also countered malicious jamming by directing a null toward the jammer, achieving effective jamming mitigation and high secrecy capacity in 2D and 3D, even with errors in locating the eavesdropper.

On the other hand, CLAs are also applied in other wireless applications. In \cite{CLA3}, a method for synthesizing shaped-beam cylindrical conformal arrays was proposed, utilizing element rotation and phase optimization. Adjustments to the vectorial active element pattern allowed rotations to be transformed and summed in a common system. Meanwhile, \cite{CLA2} explored the use of CLAs in microwave and millimeter-wave imaging for detecting concealed weapons, focusing on balancing speed and cost. A cylindrical sparse MIMO array was proposed in \cite{CLA2} to enhance speed and widen angular observations without mechanical scanning.
\begin{figure*}
	\centering 
	\includegraphics[width=6.25in]{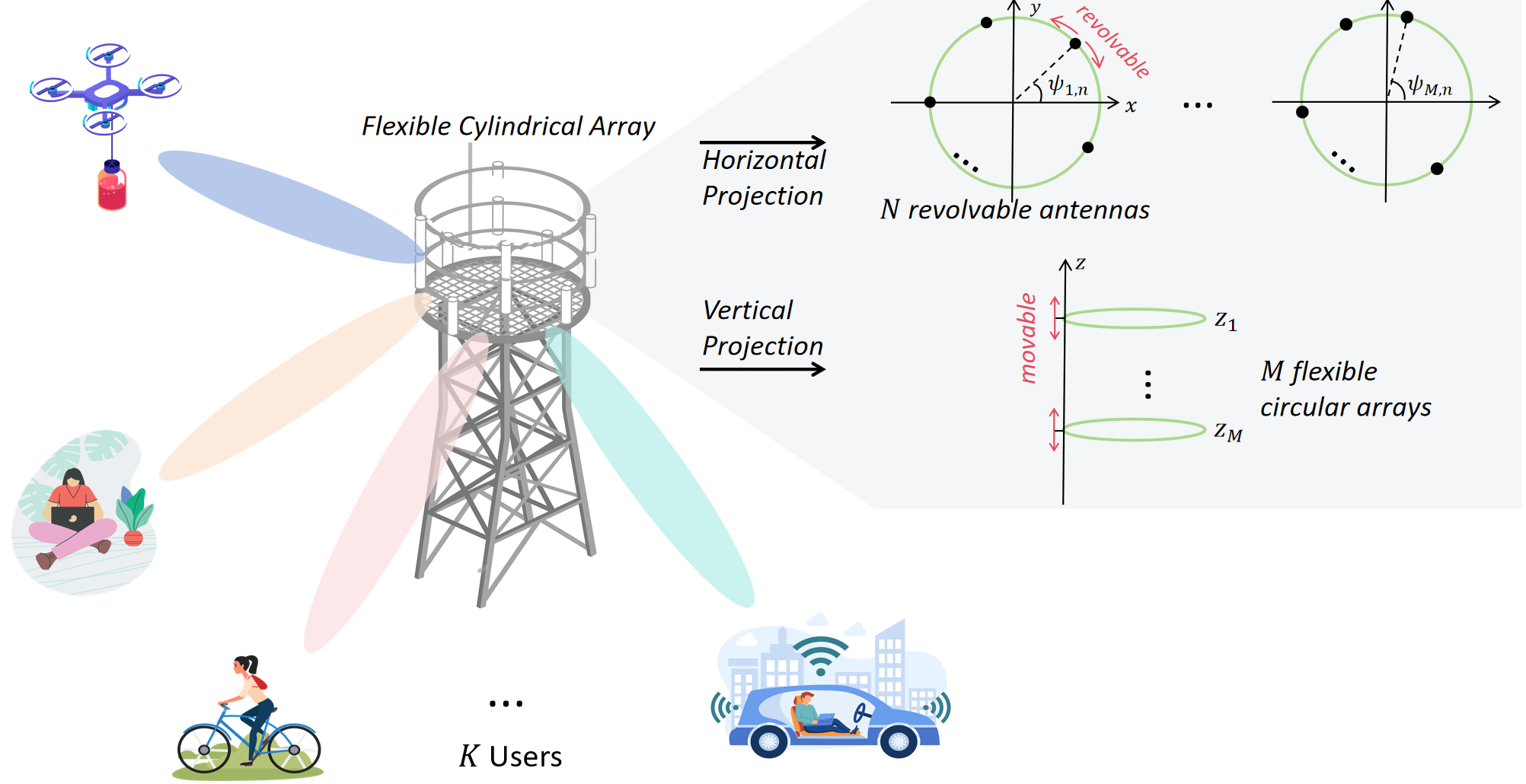}
	\caption{The multi-user communication scenario and the FCLA model.}\label{sys} 
\end{figure*}  
\subsection{Related Works and Contributions}

In dynamic scenarios, we introduce the concept of flexible cylindrical arrays (FCLAs), a novel approach that combines flexible antennas with cylindrical arrays. This concept mirrors the philosophy of flexible conformal arrays as discussed in \cite{Fcon1, Fcon2}, and bendable arrays in \cite{FAA1}, both of which emphasize the adaptable geometry of cylindrical arrays. For example, a planar array can be bent to form a semi-cylindrical array, with the arc's size dependent on the bending angle. By leveraging this dynamic geometry, specific goals of array synthesis, such as reducing side-lobes, increasing directivity, and achieving desired radiation patterns, are enhanced. This was explored in \cite{SCA1}, where array synthesis for cylindrical arrays was examined to optimize antenna coefficients and positions on a cylindrical surface.
Furthermore, our previous research \cite{P_FAA} extended the bendable array concept to wireless communication applications, evaluating the impact of bending angles on communication systems. The study found that exploiting the degrees of freedom offered by bending could improve received power, minimize the Cram\'er-Rao bound, and maximize the sum-rate by altering array elements' positions and orientations.

Building on this foundation, we propose an FCLA with a uniquely shaped geometry, consisting of several flexible circle arrays (FCAs) that can move along the vertical axis, with antenna elements capable of revolving around the circle. This geometry fully utilizes the advantages of cylindrical arrays, along with flexible antenna positioning and orientation, to enhance wireless communications. Special attention is given to the directional radiation pattern for each antenna, optimizing performance in dynamic wireless environments.
 The main contributions are summarized as follows:
\footnote{\label{note1}The source code of this work is available at \url{https://github.com/YyangSJ/flexible-cylindrical-arrays-beamforming} for readers studying.}
 
 \begin{itemize}
 	\item We model the antenna position and orientation mappings for the proposed FCLA. Specifically, the position of each antenna element lying on one FCA is represented by a revolving angle given the radius of the circle track. Then, the multipath channel is modeled with respect to the revolving angles and the vertical positions.
 	\item We utilize the regularized zero-forcing (RZF) precoding scheme 
 	 to formulate an optimization problem with respect to antenna coefficients and positions. Notably, each antenna position consists of the vertical position and the revolving angle. Hence, by using a dictionary representation for antenna position, 
 	  a customized compressive sensing (CS) framework is proposed to jointly optimize the precoding and antenna position. In this framework, three key different aspects from the standard CS framework are emphsized, namely, the regularization term, the group-wise sparsity, and the atom constraint.
 	  \item We propose to address the above CS problems with two methods. The first method aims to jointly optimize the vertical position, the revolving angle, and precoding coefficients, by iteratively selecting antenna and computing its precoding coefficients across all data streams. The second method alternatively optimizes the vertical position and the revolving angle by solving sparse subproblems. 
 \end{itemize}

The rest of this paper is organized as follows: Section \ref{sys_model} provides the signal model and the FCLA model. Section \ref{S3} formulates the CS problem to maximize the sum-rate with respect to precoding coefficients and antenna positions. Section \ref{S4} proposes two different methods to solve the optimization problem. Section \ref{simu} evaluates the proposed methods with numerical simulations. Finally, Section \ref{Con} concludes this paper.

{\emph {Notations}}:
  ${\left(  \cdot  \right)}^{ T}$, ${\left(  \cdot  \right)}^{ H}$, and $\left(\cdot\right)^{-1}$ denote   transpose, conjugate transpose, and inverse, respectively. $\vert\cdot\vert$ denotes the amplitude of a complex-value number.
$\Vert\cdot\Vert_0$ and $\Vert\cdot\Vert_1$ represent $\ell_0$ norm and $\ell_1$ norm, respectively. 
$\Vert\mathbf{A}\Vert_F$ denotes the Frobenius norm of matrix $\mathbf{A}$.     $[\mathbf{A}]_{i,:}$ and $[\mathbf{A}]_{:,j}$ denote the $i$-the row and the $j$-the column of matrix $\mathbf{A}$, respectively.  $\mathbb{E}\{\cdot\}$ denotes the expectation.

\section{System Model}\label{sys_model}

We examine a MU-MISO downlink system, as shown in Fig. \ref{sys}, where the BS features a FCLA composed of $M$ FCAs. Each FCA is equipped with $N$ antennas capable of revolving around the circular track. The revolving angle for the $n$-the element in the $m$-th FCA is denoted by $\psi_{m,n}$, $m\in\{1,\cdots,M\}$, $n\in\{1,\cdots,N\}$. Moreover, the $m$-th FCA is capable of moving along the $z$-axis, with a height of $z_m$. 
The system supports $K$ users, each with a single antenna, distributed in a full $360^\circ$ area around the BS. Furthermore, the BS provides a single data stream to each user.

In the downlink communication system, the received signal at the $k$-th user, $k\in\{1,\cdots,K\}$, can be expressed by
\begin{equation}
	y_k=\mathbf{h}_k^H\mathbf{F}\mathbf{s}+n_k,
\end{equation}
where $\mathbf{h}_k\in\mathbb{C}^{MN\times 1}$ is the $k$-th user's channel, $\mathbf{F}\triangleq[\mathbf{f}_1,\cdots,\mathbf{f}_K]\in\mathbb{C}^{MN\times K}$ is the precoding matrix,
$\mathbf{s}\in\mathbb{C}^{K\times 1}$ represents the $K$ data streams for $K$ users, $n_k$ represents the Gaussian additive white noise following $\mathcal{CN}(0,\sigma^2)$.

By assuming the i.i.d. transmit data such that $\mathbb{E}\{s_k^*s_k\}=1$ and $\mathbb{E}\{s_i^*s_k\}=0$, $\forall k , k\neq i$,
the SINR of the $k$-user can be given by
\begin{equation}\label{SINR}
	{\rm SINR}_k=\frac{\vert\mathbf{h}_k^H\mathbf{f}_k\vert^2}{\sum_{i,i\neq k}^{K}\vert\mathbf{h}_k^H\mathbf{f}_i\vert^2+\sigma^2}.
\end{equation} 

By assuming $L$ paths for all user channels, the spatial channel under the far-field plane-wave assumption is given by, $\forall k$,
 \begin{equation}\label{hk}
 	\mathbf{h}_k=\sqrt{\frac{1}{L}}\sum_{l=1}^{L}\beta_{k,l} \mathbf{a}(\vartheta_{k,l},\varphi_{k,l})\odot\mathbf{e}(\vartheta_{k,l},\varphi_{k,l}),
 \end{equation}
where $\beta_{k,l}$, $k\in\{1,\cdots,K\}$, $l\in\{1,\cdots,L\}$, is the complex path gain of the $l$-th path of the $k$-user's channel,  $\vartheta_{k,l}$ and $\varphi_{k,l}$ correspond to the elevation and azimuth angles. The array-angle manifold $\mathbf{a}\in\mathbb{C}^{MN\times 1}$ is a column-stacked vector by $\mathbf{a}\triangleq[\mathbf{a}_1^T,\cdots,\mathbf{a}_M^T]^T$, with $M$ and $N$ denoting the elevation and azimuth antenna counts, respectively, given by
\begin{equation}
	\begin{aligned}
	&	\left[\mathbf{a}_m(\vartheta,\varphi)\right]_n \\
		=& e^{-j\frac{2\pi}{\lambda}( x_{m,n}\sin\vartheta\cos\varphi + y_{m,n}\sin\vartheta\sin\varphi  +z_n\cos\vartheta) }
		\\=&e^{-j\frac{2\pi}{\lambda}( R\cos\psi_{m,n}\sin\vartheta\cos\varphi + R\sin\psi_{m,n}\sin\vartheta\sin\varphi +z_n\cos\vartheta) } \\
		=& e^{-j\frac{2\pi}{\lambda}( \phi^x R \cos\psi_{m,n}+  \phi^y R \sin\psi_{m,n}+z_m \theta) } ,
	\end{aligned}
\end{equation}
where the second equation holds due to the antenna position mapping in the circle orbit, i.e., $x_{m,n}=R\cos\psi_{m,n}$ and $y_{m,n}=R\sin\psi_{m,n}$. Moreover,
 $\phi^x\triangleq \sin\vartheta\cos\varphi $, $\phi^y\triangleq \sin\vartheta\sin\varphi $, and $\theta\triangleq \cos\vartheta$ are defined as the virtual azimuth and elevation angles for clarity. $\mathbf{e}\in\mathbb{C}^{MN\times 1}$ denotes the antenna pattern coefficient, in most studies, it is set to $\mathbf{1}$ for omni-directional radiation. However, in practical systems, directional patterns are preferred since the array could focus more energy on the area of interest.

 The cosine pattern that is widely utilized in the antenna community, particularly for representing directional radiation patterns \cite{cos1,cos2}. This pattern is useful for modeling antennas that have their main lobe oriented along the $z$-axis, such as directional or beam antennas.
 The mathematical expression of the cosine pattern for an antenna oriented along the $z$-axis is given by:
 \begin{equation}\label{rq}
 	Q_E(\vartheta,\varphi)=\begin{cases}
 		Q\cos^\kappa \vartheta, & \vartheta\in [0,\frac{\pi}{2}], \varphi \in [0,2\pi] \\
 		0, & \text{otherwise}
 	\end{cases},
 \end{equation}
 where $Q\triangleq 2(\kappa+1)$ is used for normalization.
 
To apply it in our $x$-axis orientation case, the pattern response is modified to $\sin\vartheta\cos\vartheta$.
 	This modified pattern function is presented as:
 	
 	To apply the modified pattern response in the $x$-axis orientation case, we have
 	\begin{equation}\label{GE}
 		Q_E(\vartheta,\varphi)=
 		\begin{cases}
 			Q \sin^\kappa \vartheta \cos^\kappa \varphi,& \vartheta\in[0,\pi],\varphi\in [-\frac{\pi}{2},\frac{\pi}{2}], \\
 			0, & \text{otherwise},
 		\end{cases}
 	\end{equation}
 	where $\kappa\geq 1$ is the pattern sharpness factor. A higher value of $\kappa$ indicates a radiation pattern having stronger directivity. The power normalization factor $Q$ is defined to satisfy the condition  $\int_{\Omega} Q \sin^\kappa \vartheta \cos^\kappa \varphi {\rm d} \Omega = 4\pi $, where $\Omega$ represents the spherical space. For this scenario, $Q$ is determined to be $2(1+\kappa)$, as shown in the following derivation.  
   \begin{equation}\label{GG}
  		\begin{aligned}
  			\int\int Q\sin^\kappa\vartheta \cos^\kappa \varphi {\rm d} \Omega&=\int_{-\frac{\pi}{2}}^{\frac{\pi}{2}}	\int_{0}^{\pi} Q\sin^{\kappa+1}\vartheta \cos^\kappa \varphi {\rm d} \vartheta {\rm d}\varphi \\ 
  			&\equiv  4\pi.
  		\end{aligned}
  	\end{equation}
  	The integral can be derived with the Wallis formula:
  	\begin{equation}\label{GG2}
  		\begin{aligned}
  			&\int_{-\frac{\pi}{2}}^{\frac{\pi}{2}}	\int_{0}^{\pi} Q\sin^{\kappa+1}\vartheta \cos^\kappa \varphi {\rm d} \vartheta {\rm d}\phi  \\
  			&= 4Q \int_{0}^{\frac{\pi}{2}}  \sin^{\kappa+1}\vartheta  {\rm d} \vartheta	\int_{0}^{\frac{\pi}{2}}  \cos^\kappa \varphi {\rm d}\varphi 
  			\\
  			&=4Q\cdot\frac{n}{n+1}\frac{n-2}{n-1}\cdots \frac{4}{5}\frac{2}{3} \cdot \frac{n-1}{n}\frac{n-3}{n-2}\cdots \frac{3}{4}\frac{1}{2}\frac{\pi}{2} \\
  			&=\frac{2\pi Q}{\kappa+1}.
  		\end{aligned}
  	\end{equation}
  	Therefore, Eqs. (\ref{GG}) and (\ref{GG2}) yield $Q=2(\kappa+1)$ for Eq. (\ref{GE}). 
  	
  	For the antenna pattern $\mathbf{e}(\vartheta_{k,l},\varphi_{k,l})$, we can also divide it into $\mathbf{e}\triangleq [\mathbf{e}_1^T,\cdots,\mathbf{e}_M^T]^T$. Noticing that the antenna pattern is related to the antenna
  	each entry of $\mathbf{e}_m$, $\forall m$, is denoted by
  	\begin{equation}
  	[\mathbf{e}_m(\vartheta_{k,l},\varphi_{k,l})]_n=\sqrt{Q_E(\vartheta_{k,l},\varphi_{k,l}-\psi_{m,n})}. 
  	\end{equation} 
\section{CS-Based Flexible RZF Precoding}\label{S3}
 
In FCLAs, the revolving angles and heights, denoted by $\bm{\psi}\triangleq \{\psi_{m,n}\}_{m=1,n=1}^{M,N}$ and $\mathbf{z}\triangleq \{z_m\}_{m=1}^M$  will have an impact on the channel $\mathbf{h}_k,\forall m$. Consequently, maximizing the SINR in Eq. (\ref{SINR}) depends not only on $\mathbf{F}$ but also on $\bm{\psi}$ and $\mathbf{z}$. Therefore, the
sum-rate optimization problem in FCAs is 
\begin{equation}\label{p1}
	\begin{aligned}
		&	\underset{\mathbf{F},\bm{\psi},\mathbf{z}}{\rm arg\ max} \	\sum_{k=1}^{K}\log(1+\text{SINR}_k(\mathbf{F},\bm{\psi},\mathbf{z})) \\ 
	{\rm s.t.}\	&  \Vert\mathbf{F}\Vert_F^2 \leq P,\\
		&\vert\psi_{m,i}-\psi_{m,j}\vert \geq \psi_{\rm min}, \\  & \vert z_{\overline{i}}-z_{\overline{j}}\vert \geq d_{\rm min}, \\
		&\forall m\in\{1,\cdots,M\}, \forall i,j\in\{1,\cdots,N\}, \ i\neq j,  \\
		&  \forall \overline{i},\overline{j}\in\{1,\cdots,M\}, \ \overline{i}\neq \overline{j}.
	\end{aligned}
\end{equation} 
where $P$ is the total transmit power, $\psi_{\rm min}$ is the minimal angle between arbitrary two elements within one FCA, and $d_{\rm min}$ is the minimal spacing between arbitrary two FCAs.  $\psi_{\rm min}$ and $d_{\rm min}$ are designed to ensure that the antenna elements are not positioned too closely together, thereby avoiding severe mutual coupling.

Minimizing the multi-user interference (MUI) is equivalent to maximizing the sum-rate in Eq. (\ref{p1}).
By stacking the channel vectors of all users in rows, denoted by $\mathbf{H}=[\mathbf{h}_1,\cdots,\mathbf{h}_K]^H\in\mathbb{C}^{K\times N}$, we consider the RZF precoding problem for minimizing the MUI and the transmit power:
\begin{equation}\label{FZF_pro}
	\underset{\mathbf{F}}{{\rm arg \ min}} \	\left\Vert\mathbf{I}_K-\mathbf{H}\mathbf{F}\right\Vert_F^2+\alpha\left\Vert \mathbf{F} \right\Vert_F^2.
\end{equation}
where $\alpha$ is the regularization factor. Its solution is
\begin{equation}\label{RZF}
	\mathbf{F}=\mathbf{H}^H (\mathbf{H}\mathbf{H}^H+\alpha\mathbf{I}_K )^{-1}.
\end{equation}
To satisfy the total power constraint, $\mathbf{F}$ requires further processing for normalization.

Let $\alpha$ be adjustable, general RZF can be equivalent to MRT, ZF, and MMSE precoding:
\begin{equation}\label{FFF}
\mathbf{F}=	\begin{cases}
		\mathbf{F}_{\rm MRT}\triangleq \mathbf{H}^H, \alpha\rightarrow\infty; \\
	\mathbf{F}_{\rm ZF}\triangleq \mathbf{H}^H (\mathbf{H}\mathbf{H}^H)^{-1},\alpha = 0; \\
		\mathbf{F}_{\rm MMSE}\triangleq \mathbf{H}^H (\mathbf{H}\mathbf{H}^H+\sigma^2\mathbf{I}_K )^{-1}, \alpha =\sigma^2.
	\end{cases}
\end{equation}

With the variables $\bm{\psi}$ and $\mathbf{z}$, the flexible RZF precoding problem is formulated by
\begin{equation}\label{FZF_2}
	\begin{aligned}
		\underset{\mathbf{F},\bm{\psi},\mathbf{z}}{{\rm arg \ min}}& \	\left\Vert\mathbf{I}_K-\mathbf{H}\mathbf{F}\right\Vert_F^2+\alpha\left\Vert \mathbf{F} \right\Vert_F^2\\
		 {\rm s.t.} \ & \vert\psi_{m,i}-\psi_{m,j}\vert \geq \psi_{\rm min}, \\  & \vert z_{\overline{i}}-z_{\overline{j}}\vert \geq d_{\rm min}, \\
		 &\forall m\in\{1,\cdots,M\}, \forall i,j\in\{1,\cdots,N\}, \ i\neq j,  \\
		 &  \forall \overline{i},\overline{j}\in\{1,\cdots,M\}, \ \overline{i}\neq \overline{j}.
	\end{aligned} 
\end{equation}

Since it is hard to solve the above non-convex problem, we seek for efficient frameworks to describe this problem.
Noticing that each column of $\mathbf{H}$, represents an array-position manifold (APM), depends on the antenna position,  we can re-write $\mathbf{H}$   as follows:
\begin{equation}
	\begin{aligned}
		\mathbf{H}=&[\mathbf{H}_1,\cdots,\mathbf{H}_M]
		\\ =&\left[\mathbf{b}(\psi_{1,1},z_{1}),\cdots,\mathbf{b}(\psi_{1,N},z_1), \cdots, \right. \\ & \left. \mathbf{b}(\psi_{M,1},z_{M}),\cdots,\mathbf{b}(\psi_{M,N},z_M)
		\right],
	\end{aligned}
\end{equation}
where $\mathbf{H}_m\triangleq [\mathbf{b}(\psi_{m,1},z_m),\cdots,\mathbf{b}(\psi_{m,N},z_m)]$, $\forall m$. Ignoring the subscript, the $k$-th entry of the APM is given by 
\begin{equation}
	\begin{aligned}
	[\mathbf{b}(\psi,z)]_k 
	=&\frac{1}{\sqrt{L}} 
 \sum_{l=1}^L \beta_{1,l}^* \sqrt{Q
	}\sin^{\frac{\kappa}{2}}\vartheta_{k,l}\cos^{\frac{\kappa}{2}}(\varphi_{k,l}-\psi) \\ &\ \ \ \ \ \ \ \times e^{-j\frac{2\pi}{\lambda} (R\phi^x_{k,l}\cos\psi+R\phi^y_{k,l}\sin\psi+z\theta_{k,l}) }.
	\end{aligned}
\end{equation}

Subsequently, we identify  $N$ revolving angles in each FCA in a contious space $\psi_n\in[0,2\pi]$ and determine $M$ heights along the $z$-axis range. This can be achieved using CS frameworks through dictionary representation:
\begin{equation}\label{H_dic}
	\mathbf{HF}\approx  \widetilde{\mathbf{H}}\widetilde{\mathbf{F}},
\end{equation} 
where $\widetilde{\mathbf{H}}\triangleq\left[\widetilde{\mathbf{H}}_{{\rm h},},\cdots,\widetilde{\mathbf{H}}_{{\rm h},{G_V}}\right]\in\mathbb{C}^{K\times G_VG_H}$ and $\widetilde{\mathbf{F}}\triangleq \begin{bmatrix}
	\widetilde{\mathbf{F}}_{{\rm h},1} \\ \vdots \\ \widetilde{\mathbf{F}}_{{\rm h},G_V}
\end{bmatrix}\in\mathbb{C}^{G_VG_H\times K}$.
$ 
\widetilde{\mathbf{H}}_{{\rm h},g_v}\triangleq \left[ \mathbf{b}(\psi_{g_v,1},z_{g_v}), \mathbf{b}(\psi_{g_v,2},z_{g_v}),\cdots, \mathbf{b}(\psi_{g_v,G_H},z_{g_v})
\right] \in\mathbb{C}^{K\times G_H}$ denotes the dictionary where each column represents a $(\psi,z)$ candidate.  $\widetilde{\mathbf{F}}_{{\rm h},m}\in\mathbb{C}^{G\times K}$ is a $N$-sparse matrix, meaning there are $N$ non-zero rows. These non-zero rows identify the $N$ antenna positions and their corresponding precoding coefficients. 

Considering the constraint in problem (\ref{FZF_2}), we formulate the following constrained CS problem:
\begin{equation}\label{SO}
	\begin{aligned}
		\underset{\widetilde{\mathbf{F}}}{{\rm arg \ min}} \ &	\left\Vert\mathbf{I}_K-\widetilde{\mathbf{H}}\widetilde{\mathbf{F}}\right\Vert_F^2+\alpha\left\Vert \widetilde{\mathbf{F}} \right\Vert_F^2 \\
		{\rm s.t.} 	& \left\Vert \widetilde{\mathbf{F}}\right\Vert_{0,\rm row}= MN, \\ & \vert\psi_{m,i}-\psi_{m,j}\vert \geq \psi_{\rm min}, \\  & \vert z_{\overline{i}}-z_{\overline{j}}\vert \geq d_{\rm min}, \\
		&\forall m\in\{1,\cdots,M\}, \forall i,j\in\{1,\cdots,N\}, \ i\neq j,  \\
		&  \forall \overline{i},\overline{j}\in\{1,\cdots,M\}, \ \overline{i}\neq \overline{j}, 
	\end{aligned} 
\end{equation}
where $\left\Vert\cdot\right\Vert_{0,\rm row}$ denotes the number of non-zero rows in a matrix,  
 
The above CS problem differs from standard CS problems in three key aspects:
\begin{itemize}
	\item  \emph{Regularization Term:} The regularization term $\alpha\left\Vert \widetilde{\mathbf{F}} \right\Vert_F^2$  balances the trade-off between minimizing MUI and minimizing power, helping to mitigate noise in low signal-to-noise ratio (SNR) scenarios. By varying $\alpha$, this approach can lead to three classical precoding schemes as described in Eq. (\ref{FFF}).
	\item \emph{Group-Wise Sparsity:} Note that the first constraint in Eq. (\ref{SO}) restricts the sub-matrix  $\widetilde{\mathbf{F}}_m$ to have $N$-row-sparsity, rather than imposing $NM$-row sparsity on the entire matrix $\widetilde{\mathbf{F}}$. This implies that the sparsity is organized into  $M$ groups, with each group required to exhibit $N$-row sparsity.
	\item \emph{Atom Constraint:} Observe that the second constraint in Eq. (\ref{SO}) permits the inter-element distance on each layer-circle to reach a minimal angle $\psi_{\rm min}$. Additionally, each layer-circle must satisfy the condition that the spacing exceeds $d_{\rm min}$.
\end{itemize}
  
 \section{Proposed Solutions}\label{S4}
 Recognizing the 2D sparsity characteristic of the problem in Eq. (\ref{SO}), specifically the requirement to optimize both $\mathbf{z}$ and $\bm{\psi}$, this section introduces two solutions: joint optimization and alternating optimization. We employ orthogonal matching pursuit (OMP) as the foundational algorithm, allowing for straightforward adaptation to various scenarios.

 \subsection{Joint Optimization for $\{\psi_{m,n}
 	\}_{m,n}^{M,N} $ and $\{z_{m}
 	\}_{m}^{M} $  }\label{Joint}
 
To solve the formulated CS problem in (\ref{SO}), the three challenging keys are considered in the OMP procedure. First, considering the  regularization term affected LS, we can use the RLS to replace the LS solution in traditional OMP.

In addressing group-wise sparsity, the entire support is divided into $M$ groups, each containing $N$ elements, resulting in a total of $MN$ elements. From the perspective of OMP, the goal is to select $M$ groups, each consisting of $N$ atoms, from $G_HG_V$
candidate atoms. Here, the group index indicates the vertical position  $z$, while the index within the group specifies the horizontal position $\psi$. Although $MN$ atoms are required for antenna positions, the process may initially select a number of atoms exceeding $MN$. This occurs because we cannot deterministically select which groups among the $G_V$ candidates are chosen; rather, we can only select groups that include $N$ atoms. Consequently, some atoms from other groups that do not meet the requirement of $N$ atoms might initially be selected but ultimately not included in the final support.
 For example, consider $M=1$, $N=2$, $G_V=2$, and $G_H=2$.
 This setup indicates that we should find two positions with the same vertical position in a $4$-element candidate set:  $\{(\psi_{1,1},z_1),(\psi_{1,2},z_1),(\psi_{2,1},z_2),(\psi_{2,2},z_2) \}$. In the first and second iterations, $(\psi_{1,1},z_1)$ and $(\psi_{2,1},z_2)$ are selected, respectively. Although two positions are selected, they have different vertical positions and are not located in the same plane. In the third iteration, $(\psi_{1,2},z_1)$ is selected, resulting in the group corresponding to  $z_1$ being selected. Therefore, the final selected positions are $\{(\psi_{1,1},z_1),(\psi_{1,2},z_1)\}$.
 
 To address the inter-element spacing constraint inherent in MAs, a straightforward approach within the CS framework is to construct the dictionary by utilizing the minimal spacing. This involves constructing the dictionary $\widetilde{\mathbf{H}}$ in Eq. (\ref{H_dic}) using a sampling spacing of $d_{\rm min}$ for horizontal antennas and $\psi_{\rm min}$ for elevation antennas.

The comprehensive workflow of joint optimization for $\{\psi_{m,n}
\}_{m,n}^{M,N} $ and $\{z_{m}
\}_{m}^{M} $ is detailed in Algorithm \ref{Joi}, encompassing the following steps:
\begin{itemize}

\item \emph{Atom Matching:}
Given the residual matrix $\mathbf{R} \in \mathbb{C}^{K \times K}$, which initially starts as  $\mathbf{I}_K$, our objective is to identify an atom in the candidate set $\bm{\Gamma}$ that maximizes the matching gain. Since all columns of $\mathbf{R}$ contribute to each atom matching, meaning all users play a role in optimizing each antenna position, the matching results for all columns of $\mathbf{R}$ collectively determine the final outcome. This process is mathematically represented as:
$g^\star=\underset{g\in\bm{\Gamma}}{{\rm arg \ max}} \ \left\Vert \  \left[\widetilde{\mathbf{H}} \right]_{:,g} ^{H} {\mathbf{R}}  \right\Vert_1$.
The index of the selected atom is then recorded into the support $\bm{\Lambda}$.

 \item \emph{Coefficient Calculation:}  
 According to the support $\bm{\Lambda}$, the corresponding channel matrix is denoted as $\mathbf{H}^\star=\left[ \widetilde{\mathbf{H}}\right]_{:, \bm{\Lambda}}$. Utilizing this channel matrix $\mathbf{H}^\star$, we proceed to compute the precoding coefficients for the $n$-th iteration using the RLS algorithm:
 \begin{equation}
 	\mathbf{F}^\star =\left(\mathbf{H}^{\star,H}\mathbf{H}^\star+\alpha \mathbf{I}_n\right)^{-1}\mathbf{H}^{\star,H}.
 \end{equation}
\item \emph{Residual Updating:} After obtaining the precoding based on the optimized antenna positions, the residual is determined by canceling the influence of the existing atoms. This is calculated by
$\mathbf{R}=\mathbf{I}_K-\mathbf{H}^{\star}\mathbf{F}^\star$. 
	\item \emph{Candidate Set Updating:} 
	Before proceeding with atom matching for the next iteration, it is essential to confirm and update the candidate set. Specifically, we need to verify if a particular group has reached $N$ indices. If so, the atoms in this group will not participate in subsequent atom matching processes and will be removed from the candidate set. Meanwhile, the confirmed group index is recorded in the support $\bm{\Lambda}_{\rm v}$. Lastly, it is important to check whether $M$ groups have been selected. If this condition is met, indicating that all $MN$ atoms have been selected, the iteration procedure stops.
\end{itemize}

Finally, we find the atom indices in support $\bm{\Lambda}$ satisfying that their group indices belong to the support $\bm{\Lambda}_{\rm v}$, denoted by $\bm{\Lambda}^\star$. Then, channel construction and precoding are obtained.
\begin{algorithm} 
	\caption{Joint Optimization for FCLA.} 
	\label{Joi}
	\KwData {$d_{\rm min}$, $R$, $Z$, 
		 $\alpha$.	}
	\KwResult {$\mathbf{H}^\star$ and $\mathbf{F}^\star$}
	\BlankLine
	\Begin{ 
		$\textbf{Initialization:}$
		$G_V=\left\lfloor \frac{Z}{d_{\rm min}}\right\rfloor$, 	$G_H=\left\lfloor \frac{2\pi}{\psi_{\rm min}}\right\rfloor$, $\psi_{\rm min}=2\arcsin\left(\frac{d_{\rm min}}{2R}\right)$,
		 $\bm{\Gamma}=\{1,\cdots,G_VG_H\}$, $C_m=0$, $\forall m$,  $\mathbf{R}=\mathbf{I}_K$.
		Construct dictionary $\widetilde{\mathbf{H}}$ described below Eq. (\ref{H_dic}).\\
		\For{$n=1,\cdots,G_VG_H$}{ 
			\tcp{\emph{Atom Matching}} 	  $g^\star=\underset{g\in\bm{\Gamma}}{{\rm arg \ max}} \ \left\Vert \  \left[\widetilde{\mathbf{H}} \right]_{:,g} ^{H} {\mathbf{R}}  \right\Vert_2^2$ \; 
			$\bm{\Lambda}\leftarrow \bm{\Lambda}\cup g^\star$\; \tcp{\emph{Coefficient Calculation}} 
			$\mathbf{H}^\star=\left[ \widetilde{\mathbf{H}}\right]_{:, \bm{\Lambda}}$ \;
			$\mathbf{F}^\star=\left(\mathbf{H}^{\star,H}\mathbf{H}^\star+\alpha \mathbf{I}_n\right)^{-1}\mathbf{H}^{\star,H}$ \;	\tcp{\emph{Residual Updating}} 	 	$\mathbf{R}=\mathbf{I}_K-\mathbf{H}^\star\mathbf{F}^\star$\;
			\tcp{\emph{Candidate Set Updating}} 	
			\For{$m=1,\cdots,M$}{\If{$\left\lceil\frac{g^\star}{G}\right\rceil=m $}{$C_m=C_m+1$\;} \If{$C_m=N$}{
					$\bm{\Lambda}_{\rm v}\leftarrow\bm{\Lambda}_{\rm v}\cup m$\;
					$\bm{\Gamma}\leftarrow \bm{\Gamma}  \setminus \left\{
					(m-1)G_H+g_h| g_h=1,\cdots,G_V\right\} $\;
					$\overline{C}=\overline{C}+1$\;}}
				\If{$\overline{C}=M$}{ 
					Break; 
			 }  }
			 	Find all index $g$ in $\bm{\Lambda}$  satisfying that $\lceil \frac{g}{G_H}\rceil$ can be found in $\bm{\Lambda_{\rm v}}$, and record into $\bm{\Lambda}^\star$
			 	\;
			 		$\mathbf{H}^\star=\left[ \widetilde{\mathbf{H}}\right]_{:, \bm{\Lambda}^\star}$ \;
			 	$	\mathbf{F}^\star=\mathbf{H}^{\star,H} (\mathbf{H}^{\star}\mathbf{H}^{\star,H}+\alpha\mathbf{I}_K )^{-1}$ \;
		Normalize each column of $\mathbf{F}^\star$.
	}	\Return{$\mathbf{H}^\star$ and $\mathbf{F}^\star$}
\end{algorithm}

 \subsection{Optimization for $\{\psi_{m,n}
 	\}_{m,n}^{M,N} $ and $\{z_{m}
 	\}_{m}^{M} $ }\label{Alter}
Due to the extensive search range for $\{\psi, z\}$, the joint optimization proposed in Section \ref{Joint} is associated with high complexity. To effectively address this issue, we propose an alternating optimization approach in this section. By considering the sparse structure of the antenna distribution, it is more efficient to optimize the height by treating each FCA as a whole. Additionally, we can optimize revolving angles within different FCAs in parallel, allowing for a reduced number of iterations.

\subsubsection{Revolving Angle Optimization}

Initially, the heights for all FCAs are generated and fixed as $\{z_1,\cdots,z_M\}$ during the optimization of revolving angles. In this context, there is no need to set candidate for antenna heights, which subsequently reduces the size of the virtual channel. Consequently, the dictionary representation can be expressed as follows:
\begin{equation}\label{H_dic2} 
	\mathbf{HF}\approx  \widetilde{\mathbf{H}}_{\rm h}\widetilde{\mathbf{F}}_{\rm h},
\end{equation} 
where $\widetilde{\mathbf{H}}_{\rm h}\triangleq\left[\widetilde{\mathbf{H}}_{{\rm h},1},\cdots,\widetilde{\mathbf{H}}_{{\rm h},M}\right]\in\mathbb{C}^{K\times MG_H}$,
\begin{equation}\label{Hhm}
	\widetilde{\mathbf{H}}_{{\rm h},m}\triangleq \left[ \mathbf{b}(\psi_{m,1},z_{m}), \mathbf{b}(\psi_{m,2},z_{m}),\cdots, \mathbf{b}(\psi_{m,G_H},z_{m})
	\right],
\end{equation} and $\widetilde{\mathbf{F}}_{\rm h}\triangleq \begin{bmatrix}
	\widetilde{\mathbf{F}}_{{\rm h},1} \\ \vdots \\ \widetilde{\mathbf{F}}_{{\rm h},M}
\end{bmatrix}\in\mathbb{C}^{MG_{\rm h}\times K}$. The above representation resembles Eq. $\ref{H_dic}$, with the distinction that the heights of the $M$ FCAs are fixed.

\begin{algorithm} 
	\caption{Alternating Optimization for FCLA.} 
	\label{Alt}
	\KwData {$d_{\rm min}$, $R$, $Z$, 
		$\alpha$.	}
	\KwResult {$\mathbf{H}^\star$ and $\mathbf{F}^\star$}
	\BlankLine
	\Begin{ 
		$\textbf{Initialization:}$ 	$G_V=\left\lfloor \frac{Z}{d_{\rm min}}\right\rfloor$, 	$G_H=\left\lfloor \frac{2\pi}{\psi_{\rm min}}\right\rfloor$, $\psi_{\rm min}=2\arcsin\left(\frac{d_{\rm min}}{2R}\right)$,
		$\bm{\Gamma}=\{1,\cdots,G_VG_H\}$, $C_m=0$, $\forall m$,  $\mathbf{R}=\mathbf{I}_K$.
	\\
		\For{$i=1,\cdots,I$}{
				Construct $\widetilde{\mathbf{H}}_{{\rm h},m} $ according to Eqs. (\ref{Hhm})\; 	$\mathbf{R}=\mathbf{I}_K$, $\mathbf{H}^\star=[\ ]$, 	$\bm{\Lambda}_{{\rm h},m}=\emptyset$
				\;
		\For{$n=1,\cdots,N$}{   \For{$m=1,\cdots,M$}{ $g_{h,m}^\star=\underset{g_{h,m}\in\bm{\Gamma}_{\rm h}}{{\rm arg \ max}} \ \left\Vert \  \widetilde{\mathbf{H}}_{{\rm h},m,g_{h,m}}  ^{H} {\mathbf{R}}  \right\Vert_2^2$\;
				$\bm{\Lambda}_{{\rm h},m}\leftarrow \bm{\Lambda}_{{\rm h},m}\cup g_{h,m}^\star$\;
				$\mathbf{H}^\star=\left[\mathbf{H}^\star,
				\ \widetilde{\mathbf{H}}_{{\rm h},m,g_{h,m}^\star}\right]$
			}    
			$\mathbf{F}^\star=\left(\mathbf{H}^{\star,H}\mathbf{H}^\star+\alpha \mathbf{I}_{nM}\right)^{-1}\mathbf{H}^{\star,H}$ \; 	$\mathbf{R}=\mathbf{I}_K-\mathbf{H}^\star\mathbf{F}^\star$\; 
		}
		$\{\bm{\Lambda}_{{\rm h},m}\}_{m=1}^M \Rightarrow \bm{\psi}^\star$\;
			Construct $\widetilde{\mathbf{H}}_{{\rm v},m} $ according to Eqs. (\ref{Hvm}) and (\ref{Hvmg})\;
		$\mathbf{R}=\mathbf{I}_K$, $\mathbf{H}^\star=[\ ]$, 	$\bm{\Lambda}_{\rm v}=\emptyset$
		\;
		\For{$m=1,\cdots,M$}{ $g_{v}^\star=\underset{g_{v}\in\bm{\Gamma}_{\rm v}}{{\rm arg \ max}} \ \left\Vert \  \widetilde{\mathbf{H}}_{{\rm v},m,g_{m,v}}  ^{H} {\mathbf{R}}  \right\Vert_F^2$\;
			$\bm{\Lambda}_{\rm v}\leftarrow \bm{\Lambda}_{\rm v}\cup g_{m,v}^\star$\;
			$\mathbf{H}^\star=\left[\mathbf{H}^\star,
			\  \widetilde{\mathbf{H}}_{{\rm v},m,g_{m,v}^\star} 
			\right]$ \;	$\mathbf{F}^\star=\left(\mathbf{H}^{\star,H}\mathbf{H}^\star+\alpha \mathbf{I}_{mN}\right)^{-1}\mathbf{H}^{\star,H}$ \; 	$\mathbf{R}=\mathbf{I}_K-\mathbf{H}^\star\mathbf{F}^\star$\; 
			}
		$\bm{\Lambda}_{\rm v}\Rightarrow \mathbf{z}^\star$
		  }
		Normalize each column of $\mathbf{F}^\star$.
	}	\Return{$\mathbf{H}^\star$ and $\mathbf{F}^\star$}
\end{algorithm}
 
 Then, the formulated CS problem to optimize revolving angles can be given by
 \begin{equation}\label{SO1}
 	\begin{aligned}
 		\underset{\widetilde{\mathbf{F}}_{\rm h}}{{\rm arg \ min}} \ &	\left\Vert\mathbf{I}_K-\widetilde{\mathbf{H}}_{\rm h}\widetilde{\mathbf{F}}_{\rm h}\right\Vert_F^2+\alpha\left\Vert \widetilde{\mathbf{F}}_{\rm h} \right\Vert_F^2 \\
 		{\rm s.t.} 	& \left\Vert \widetilde{\mathbf{F}}_{{\rm h},m}\right\Vert_{0,\rm row}= N, \\ & \vert\psi_{m,i}-\psi_{m,j}\vert \geq \psi_{\rm min}, \\
 		&\forall m\in\{1,\cdots,M\}, \forall i,j\in\{1,\cdots,N\}, \ i\neq j.
 	\end{aligned} 
 \end{equation}

 In the aforementioned problem, we observe that the $MN$ non-zero elements in $\widetilde{\mathbf{F}}$ are not distributed arbitrarily, as each FCA must contain $N$ elements. Consequently, the constraint $\left\Vert \widetilde{\mathbf{F}}_{{\rm h},m}\right\Vert_{0,\rm row}= N$, $\forall m$, is imposed. This implies that $N$ columns must be selected in $\widetilde{\mathbf{H}}_{{\rm h},m}, \forall m$. Furthermore, the optimization of the revolving angle within each FCA can be considered independently without affecting the others. This allows for parallel atom matching across the $M$ FCAs, which collectively determine the residual signal for the subsequent iteration. Implementing parallel atom matching can potentially accelerate the algorithm in practice. Specifically, all FCAs perform atom matching in each iteration $n$, as outlined in lines 5--9 of Algorithm \ref{Alt}, followed by a joint computation of coefficients and residual calculation using the selected atoms.

 \subsubsection{Height Optimization}
 After optimizing revolving angles for all FCAs, denoted as $\bm{\psi}^\star=\{\psi_{1,1}^\star,\cdots,\psi_{M,N}^\star\}$, the heights for the FCAs can also be optimized using an OMP approach. To facilitate this, we construct the dictionary for the $m$-th FCA as follows:
 \begin{equation}\label{Hvm}
\widetilde{\mathbf{H}}_{{\rm v},m}\triangleq\left[\widetilde{\mathbf{H}}_{{\rm v},m,1},\cdots,\widetilde{\mathbf{H}}_{{\rm v},m,G_V}\right]\in\mathbb{C}^{K\times NG_V}
 \end{equation}   with 
 \begin{equation}\label{Hvmg}
 \widetilde{\mathbf{H}}_{{\rm v},m,g_v}\triangleq
 [\mathbf{b}(\psi_{m,1}^\star,z_{g_v}),\cdots,\mathbf{b}(\psi_{m,N}^\star,z_{g_v})]\in\mathbb{C}^{K\times N}.
 \end{equation} 

For the $m$-th FCA's block dictionary $\widetilde{\mathbf{H}}_{{\rm v},m}$, the corresponding coefficient matrix $\widetilde{\mathbf{F}}_{{\rm v},m}\triangleq [\widetilde{\mathbf{F}}_{{\rm v},m,1},\cdots,\widetilde{\mathbf{F}}_{{\rm v},m,G_V}]\in\mathbb{C}^{NG_V\times K}$ should exhibit a one-block row-sparsity pattern. Consequently, we have
\begin{equation}
	\begin{aligned}
		\widetilde{\mathbf{H}}_{{\rm v},m}\widetilde{\mathbf{F}}_{{\rm v},m}&= \sum_{g_v=1}^{G_V} \widetilde{\mathbf{H}}_{{\rm v},m,g_v}\widetilde{\mathbf{F}}_{{\rm v},m,g_v}
		\\
	&=\widetilde{\mathbf{H}}_{{\rm v},m,g_v^\star}\widetilde{\mathbf{F}}_{{\rm v},m,g_v^\star},
	\end{aligned}
\end{equation}
 where $g_{m,v}^\star$ denotes the index of the optimal height of the $m$-th FCA.
 In this sense, $\widetilde{\mathbf{H}}_{{\rm v},m,g_{m,v}^\star}$ and $\widetilde{\mathbf{F}}_{{\rm v},m,g_{m,v}^\star}$ represent the optimized channel matrix and precoding coefficient for the $m$-th FCA.
 
 Stacking $M$ FCAs' dictionaries yields
 \begin{equation}
 	\widetilde{\mathbf{H}}_{\rm v}\triangleq \left[\widetilde{\mathbf{H}}_{{\rm v},1},\cdots,\widetilde{\mathbf{H}}_{{\rm v},M}\right]\in\mathbb{C}^{K\times NMG_V},
 \end{equation} 
 and the corresponding sparse coefficient matrix
 \begin{equation}
\widetilde{\mathbf{F}}_{\rm v}\triangleq \left[
\widetilde{\mathbf{F}}_{{\rm v},1},\cdots,\widetilde{\mathbf{F}}_{{\rm v},M}\right]\in\mathbb{C}^{NMG_V\times K}.
 \end{equation}
 Subsequently, the following CS problem is formulated:
  \begin{equation}\label{SO2}
 	\begin{aligned}
 		\underset{\widetilde{\mathbf{F}}_{\rm v}}{{\rm arg \ min}} \ &	\left\Vert\mathbf{I}_K-\widetilde{\mathbf{H}}_{\rm v}\widetilde{\mathbf{F}}_{\rm v}\right\Vert_F^2+\alpha\left\Vert \widetilde{\mathbf{F}}_{\rm v} \right\Vert_F^2 \\
 		{\rm s.t.} 	& \left\Vert \mathbf{U} \widetilde{\mathbf{F}}_{{\rm v},m}\right\Vert_{0,\rm row}= 1, \forall m, \\  & \vert z_{\overline{i}}-z_{\overline{j}}\vert \geq d_{\rm min}, \\
 		&   \forall \overline{i},\overline{j}\in\{1,\cdots,M\}, \ \overline{i}\neq \overline{j}, 
 	\end{aligned} 
 \end{equation}
 where $\mathbf{U}\triangleq \begin{bmatrix}
 	\mathbf{1} & & \\ &\ddots & \\ & & \mathbf{1}
 \end{bmatrix}\in\mathbb{C}^{G_V\times NG_V}$ with $\mathbf{1}$ representing an $N$-dimensional all-ones row vector.

We solve this problem by finding $M$ best $g_{m,v}^\star$ values, denoted by $\{g_{m,v}^\star|m=1,\cdots,M\}$,
 to characterize $\widetilde{\mathbf{H}}_{\rm v}\widetilde{\mathbf{F}}_{\rm v}$ as
 \begin{equation}
 \widetilde{\mathbf{H}}_{\rm v}\widetilde{\mathbf{F}}_{\rm v}=\sum_{m=1}^{M}\widetilde{\mathbf{H}}_{{\rm v},m,g_{m,v}^\star}\widetilde{\mathbf{F}}_{{\rm v},m,g_{m,v}^\star}.
 \end{equation}
  The sparsity constraint $\left\Vert \mathbf{U} \widetilde{\mathbf{F}}_{{\rm v},m}\right\Vert_{0,\rm row}= 1$ in problem (\ref{SO2}) indicates that each of the $M$ FCAs can independently determine their optimal height index. However, collectively, they should aim to minimize the objective function. From the perspective of OMP, each FCA individually performs atom matching and selection, followed by a joint calculation of the RLS and updating the residual. The entire algorithm is detailed in lines 18--24 of Algorithm \ref{Alter}.

 \subsection{Algorithm Summary and Complexity Analysis}

The two algorithms discussed in Sections \ref{Joint} and \ref{Alter} effectively tackle the problem stated in (\ref{SO}), focusing on optimizing both the precoding coefficients and the array geometry. The array geometry pertains to the positions and orientations of the antennas, which are determined by their revolving angles and heights. Although both algorithms address joint optimization of the precoding coefficients and array geometry, they differ significantly in their strategies for optimizing array geometry.

 In Algorithm \ref{Joi}, we optimize the array geometry by simultaneously considering the height and revolving angle within a single dictionary, where each atom is represented by $\mathbf{b}(\psi,z)$. In this framework, $\{(\psi_{m,n},z_m)| m=1,\cdots,M, n=1,\cdots,N\}$ can be jointly optimized by selecting $MN$ atoms from a predefined set of $G_VG_H$ candidates to construct the channel. Although $\psi_{m,n}$ and $z_m$ 
 are independent, all $N$ elements within the $m$-th FCA must have the same height $z_m$. This requirement means that the desired $MN$ atoms are not distributed arbitrarily but must adhere to group-wise sparsity constraints. Traditional OMP struggles to directly match $MN$ atoms while maintaining group-wise sparsity. Thus, Algorithm \ref{Joi} addresses this by employing an over-matching strategy, conducting $G_VG_H$
 iterations instead of $MN$ iterations to ensure that the group-wise sparsity is satisfied.   The complexity of Algorithm \ref{Joi} is mainly dominated by the atom matching and RLS steps. For atom matching, a complexity of $\mathcal{O}(K^2G_VG_H)$ is required to calculate a matrix-matrix multiplication per iteration. For RLS, the inverse operator incurs in a complexity of $\mathcal{O}(n^3)$ in the $n$-th iteration. Note that $\left(\mathbf{H}^{\star,H}\mathbf{H}^\star+\alpha \mathbf{I}_n\right)^{-1}\mathbf{H}^{\star,H}\equiv \mathbf{H}^{\star,H} \left(\mathbf{H}^\star\mathbf{H}^{\star,H}+\alpha \mathbf{I}_K\right)^{-1}$, hence the complexity of RLS can be summarized by $\mathcal{O}({\rm min}\{n^3,K^3\})$. This is smaller than the complexity of atom matching since $K\ll G_VG_H$. Moreover, regarding Algorithm \ref{Joi}, the number of iterations is expected to be within the range $[MN, G_VG_H]$. Therefore, the complexity of Algorithm \ref{Joi} in the best-case scenario is $\mathcal{O}(K^2MNG_VG_H)$,  while in the worst-case scenario, it is $\mathcal{O}(K^2G_V^2G_H^2)$.
 
 In Algorithm \ref{Alt}, alternating optimization is used for adjusting revolving angles and heights, requiring the solution of two particular CS problems. The complexity analysis depends heavily on the two atom matching steps found in lines 8 and 19, which have complexities of  $\mathcal{O}(K^2MG_H)$ and $\mathcal{O}(K^2NG_V)$, respectively. Taking into account both the inner and outer loops, the overall complexity of Algorithm \ref{Alt} can be expressed as $\mathcal{O}(IK^2MN(G_H+G_V))$. This complexity is significantly reduced compared to that of Algorithm \ref{Joi} because the number of outer iterations, denoted by $I$, is usually small.

\section{Simulation Results}\label{simu}

\begin{figure}
	\centering
	\subfigure[Omni-directional pattern.]{
		\includegraphics[width=3in]{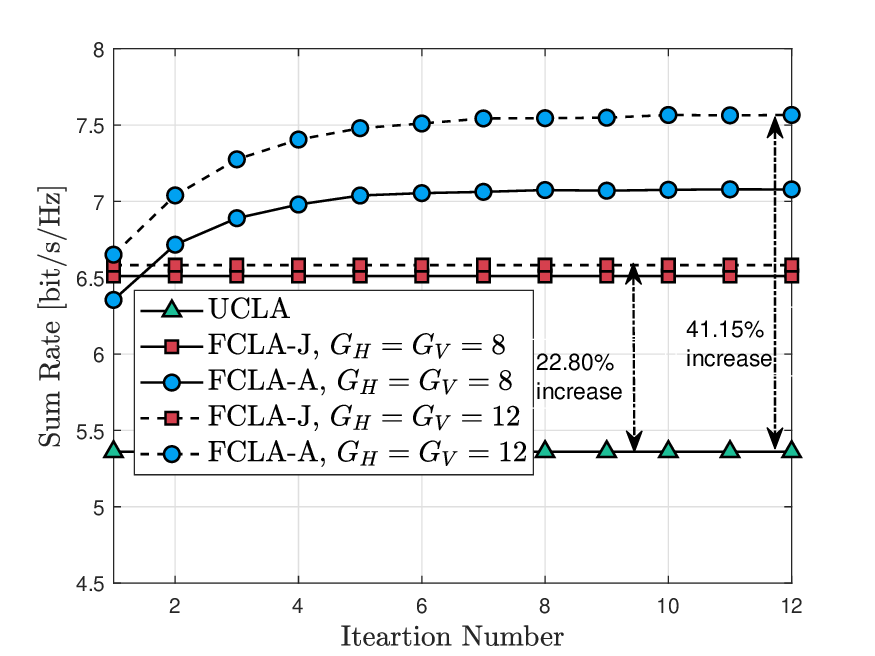}
		%\label{label_for_cross_ref_1}
	}
	\\    
	\subfigure[Directional pattern, $\kappa=1$.]{
		\includegraphics[width=3in]{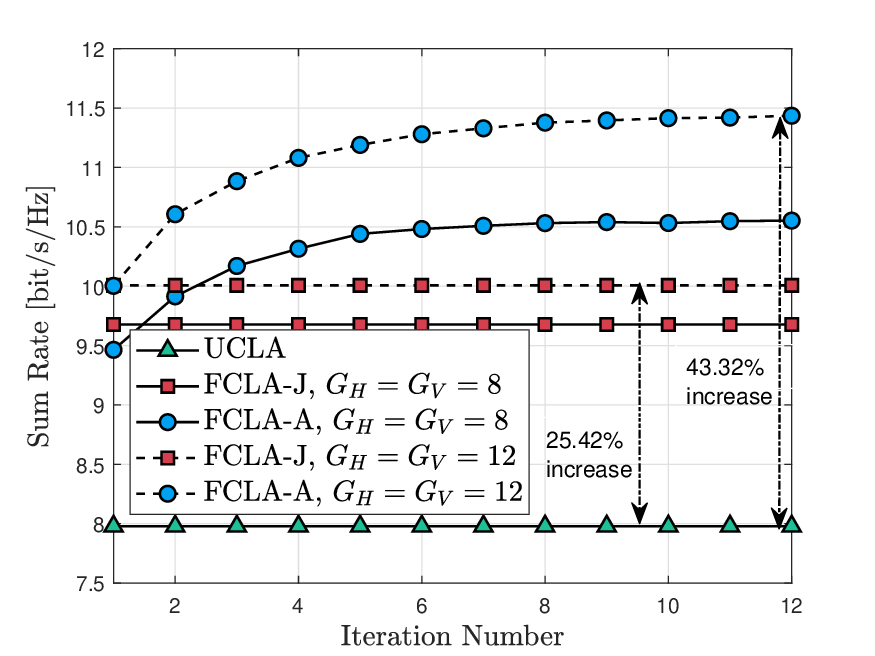}
		%	\label{label_for_cross_ref_3}
	} 
	\caption{The sum rate of different methods with omni-directional and directional patterns.}
	\label{iterk}
\end{figure}
The system operates at a central frequency of $3$ GHz. The BS is equipped with $M=4$ FCAs and each equipped with $N=4$ antennas, serving $K=16$ users.  The user and scatterer locations are uniformly distributed with azimuth angle $\varphi_{k,l}$ ranging within $ [0,2\pi]$ and elevation angle $\vartheta_{k,l}$ ranging within $[\frac{\pi}{6},\frac{5\pi}{6}]$ for all $k,l$. All users are assumed to have an identical number of channel paths $L=4$. The noise power is set to $1$, and the SNR is defined by $P$.
 The methods included in the evaluation are listed as follow.  
\begin{itemize}
	\item \textbf{UCLA:} Using RZF precoding with the uniform cylindrical array (UCLA).
	
	\item \textbf{FCLA-J:} Jointly optimizing RZF precoding and antenna positions, where antenna revolving angles  and antenna heights are jointly optimized, as shown in Algorithm \ref{Joi}.
	
	\item \textbf{FCLA-A:} Jointly optimizing RZF precoding and antenna positions, where antenna revolving angles  and antenna heights are alternatingly optimized, as shown in Algorithm \ref{Alt}.
\end{itemize}

As illustrated in Algorithm \ref{Alt}, the FCLA-A approach is dependent on the number of iterations $I$, and we aim to demonstrate the convergence of FCLA-A through numerical results. Fig. \ref{iterk} presents the performance curves of the three approaches in terms of the sum rate, where the SNR is set to 0 dB, the movable region is configured with two values $ G_H = G_V = 8$ and $G_H = G_V = 12$, and the antenna radiation pattern includes two types: the widely used omni-directional pattern and the directional pattern described in Eq. (\ref{GE}) with $ \kappa = 1$. From Fig. \ref{iterk}, it can be observed that FCLA-A converges effectively as the number of iterations increases, with convergence beginning at approximately 5 iterations. Notably, UCLA and FCLA-J exhibit no performance changes since they are independent of the number of iterations $I$.
Fig. \ref{iterk}(a) and (b) reveal that FCLA-A achieves a larger performance gain compared to FCLA-J upon convergence. This is attributed to the group-wise sparsity constraint affecting FCLA-J, as discussed in Section \ref{S3}. In comparison to UCLA, both FCLA-J and FCLA-A demonstrate significant performance improvements. For instance, in Fig. \ref{iterk}(a), FCLA-A and FCLA-J with $G_H = G_V = 12$ achieve performance gains of   41.15\%  and   22.80\%, respectively. Similarly, in Fig. \ref{iterk}(b), FCLA-A and FCLA-J with $G_H = G_V = 12$ achieve performance gains of  43.32\% and 25.42\%, respectively.
A comparison between Fig. \ref{iterk}(a) and (b) indicates that all methods exhibit performance improvements when using the directional pattern with $\kappa = 1$. For example, the sum rate of UCLA increases from approximately  5.35 to  7.95, representing an improvement of about 48.6\% when adopting the directional pattern with $\kappa = 1$. Similarly, FCLA-A with $G_H = G_V = 12$ increases from approximately  7.55  to  11.45, reflecting an improvement of about 47.7\% under the same conditions. These results underscore the significant impact of the directional pattern on the performance of CLAs.

\begin{figure}
  	\centering
  	\subfigure[Omni-directional pattern.]{
  		\includegraphics[width=3in]{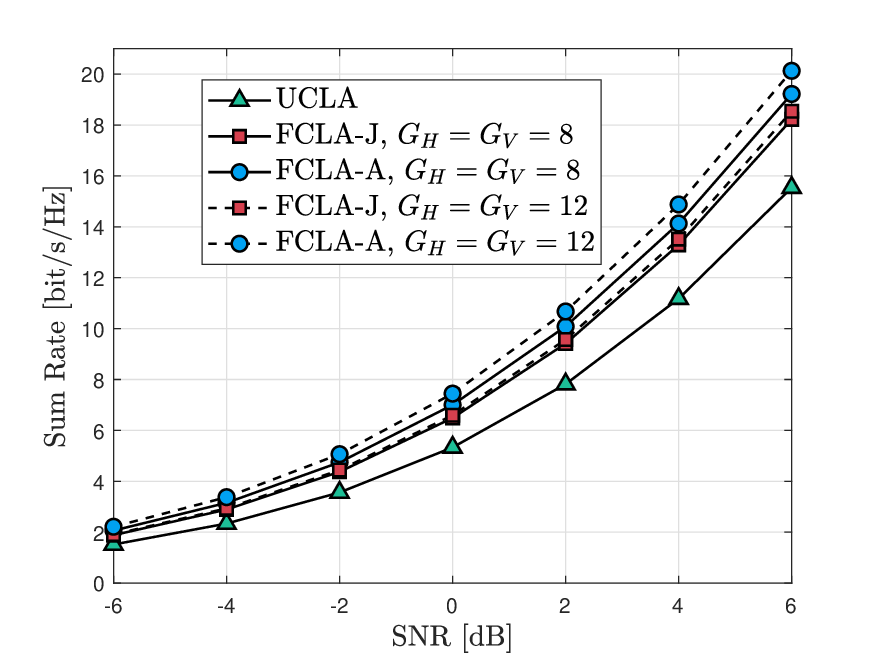}
  		%\label{label_for_cross_ref_1}
  	}
  	\\    
  	\subfigure[Directional pattern, $\kappa=1$.]{
  		\includegraphics[width=3in]{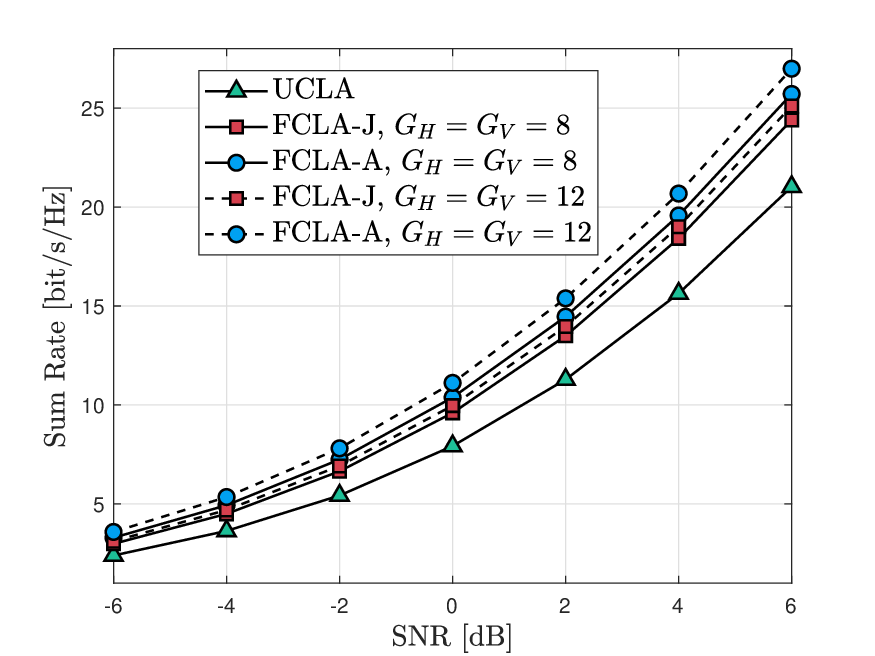}
  		%	\label{label_for_cross_ref_3}
  	} 
  	\caption{The sum rate of different methods with omni-directional and directional patterns.}
  	\label{snr}
  \end{figure}

 As depicted in Fig. \ref{snr}, we examine the impact of SNR on the sum rate across different scenarios. The SNR is varied from -6 to 6 dB in increments of 2 dB, with the iteration number $I = 5$ for FCLA-A, and the size of the movable regions set to two configurations: $G_H = G_V = 8$ and $G_H = G_V = 12$. From Fig. \ref{snr}, it is observed that the sum rate increases as the SNR rises. Consistent with the findings in Fig. \ref{iterk}, both FCLA-J and FCLA-A achieve significant performance gains compared to UCLA, with FCLA-A outperforming FCLA-J. Furthermore, a comparison between Fig. \ref{snr}(a) and (b) reveals that the directional pattern with $kappa = 1$ provides a performance advantage over the omni-directional pattern.

\begin{figure}
  	\centering
  	\subfigure[Omni-directional pattern.]{
  		\includegraphics[width=3in]{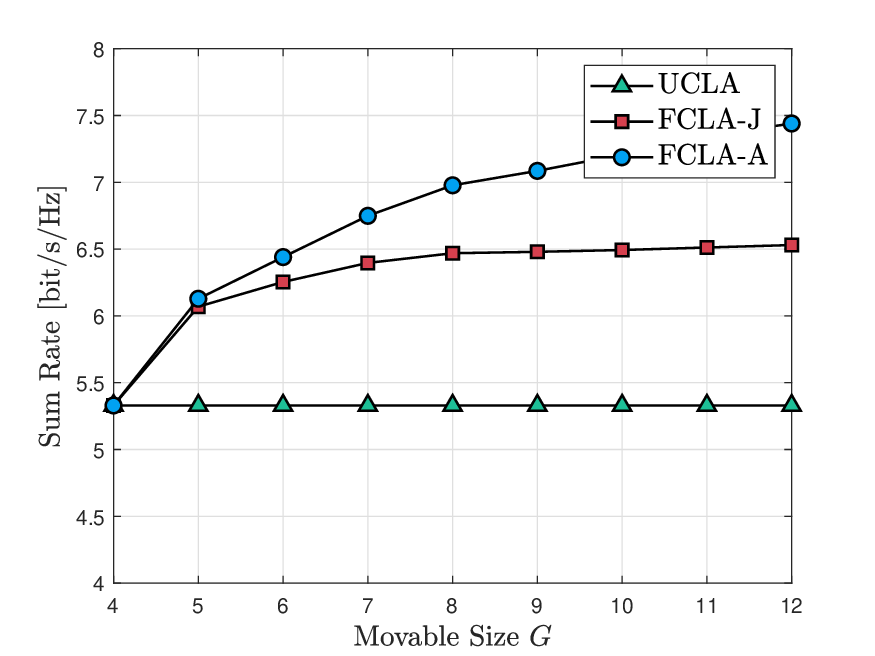}
  		%\label{label_for_cross_ref_1}
  	}
  	\\    
  	\subfigure[Directional pattern, $\kappa=1$.]{
  		\includegraphics[width=3in]{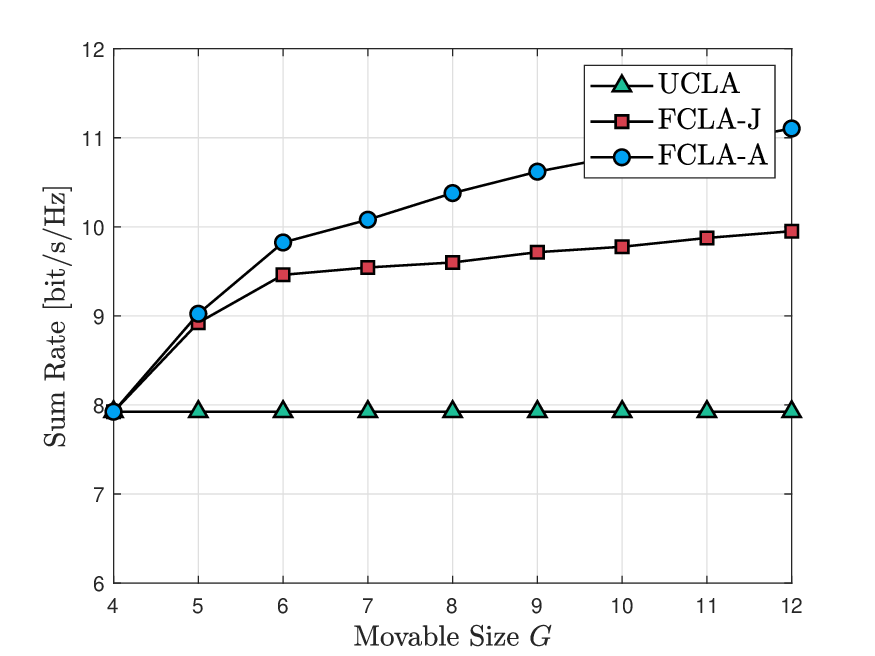}
  		%	\label{label_for_cross_ref_3}
  	} 
  	\caption{The sum rate of different methods with omni-directional and directional patterns. SNR is set to 0 dB and the number of iterations $I$ for FCLA-A is set to $5$.}
  	\label{mr0}
  \end{figure}

\begin{figure}
  	\centering
  	\subfigure[Omni-directional pattern.]{
  		\includegraphics[width=3in]{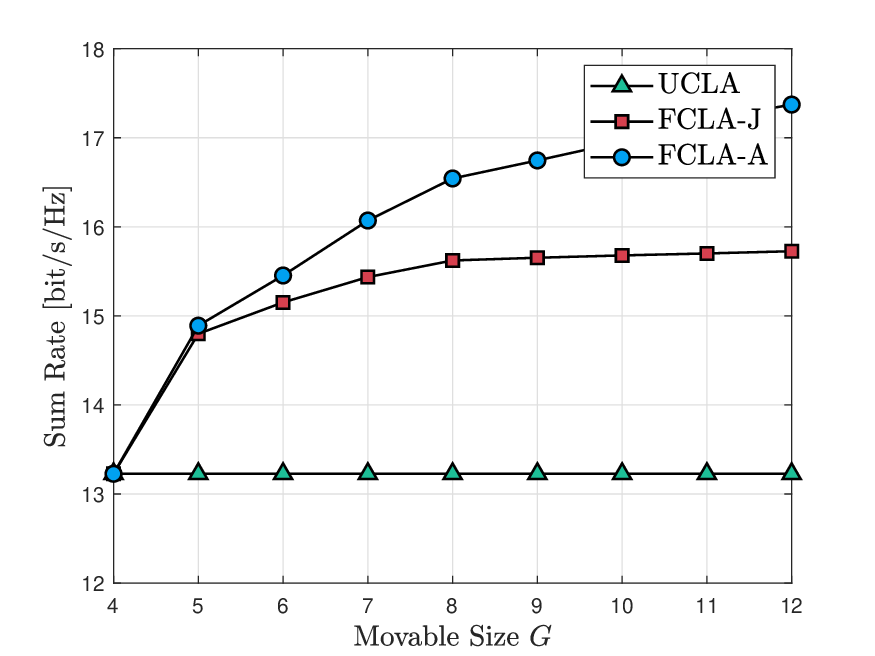}
  		%\label{label_for_cross_ref_1}
  	}
  	\\    
  	\subfigure[Directional pattern, $\kappa=1$.]{
  		\includegraphics[width=3in]{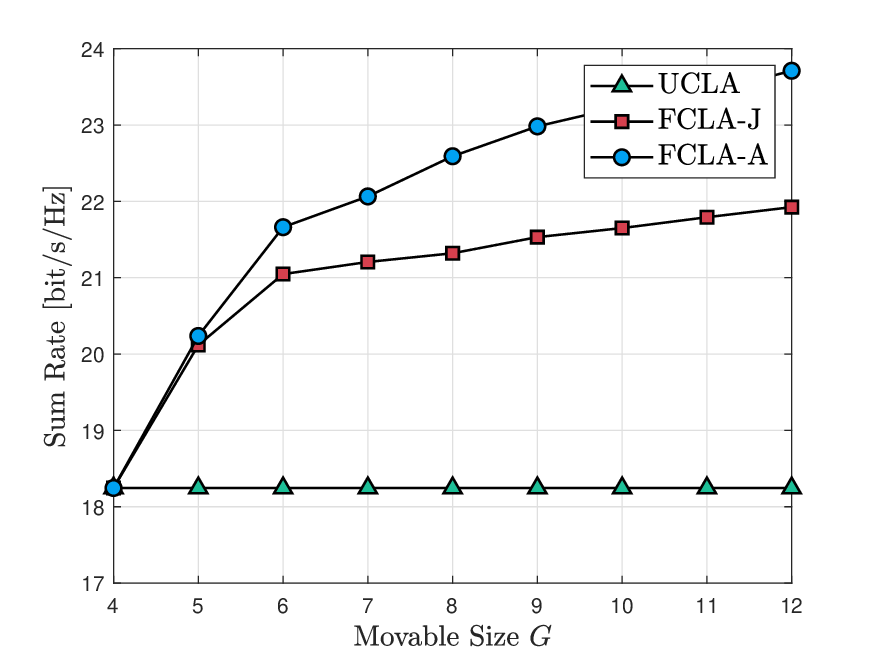}
  		%	\label{label_for_cross_ref_3}
  	} 
  	\caption{The sum rate of different methods with omni-directional and directional patterns. SNR is set to 5 dB and the number of iterations $I$ for FCLA-A is set to $5$.}
  	\label{mr5}
  \end{figure}

As depicted in Figs. \ref{mr0} and \ref{mr5}, we examine the influence of the movable region size $G = G_H = G_V$ on the sum rate across different methods, where $G$ ranges from  4  to 12), the iteration number $I = 5$ for FCLA-A, and the SNR is set to 0 dB for Fig. \ref{mr0} and 5 dB for Fig. ref{mr5}, respectively. The figures reveal that as $G$ increases, the sum rate of both FCLA-J and FCLA-A improves, but the rate of improvement slows down as $G$ becomes larger. For instance, under the omni-directional antenna pattern, FCLA-J begins to converge when $G = 10$, suggesting that the sum rate cannot increase indefinitely as the movable region expands. With directional patterns, the convergence point for the FCLA-J method shifts to a larger $G$, indicating that directional patterns offer greater potential for performance enhancement compared to omni-directional patterns.
  
\section{Conclusions}\label{Con}

 This paper explores the potential of FCLA to enhance the DoF in wireless communication systems. Building on the established FCLA model, we evaluate the sum rate in a multi-user scenario, considering both omni-directional and directional radiation patterns for the FCLA elements. To jointly optimize antenna positions (determined by revolving angles and heights) and antenna coefficients (associated with precoding), we formulate a specific CS problem based on the RZF precoding scheme. Subsequently, we propose two methods to solve this problem: FCLA-J and FCLA-A. Simulation results demonstrate that, when employing directional radiation patterns, FCLA-A and FCLA-J achieve significant performance gains of 43.32\% and 25.42\%, respectively, compared to UCLAs with RZF precoding. Additionally, FCLA-A exhibits lower computational complexity than FCLA-J.
\begin{appendices} 
\end{appendices}

\bibliographystyle{IEEEtran}
\bibliography{reference.bib}

\vspace{12pt}

\end{document}